\documentclass[latterpaper,12pt]{article}
\pdfoutput=1

\usepackage{amsmath}
\usepackage{amsfonts}
\usepackage{amssymb}
\usepackage{graphicx}
\usepackage{color}
\usepackage{slashed}
\usepackage{dsfont}
\usepackage{bbold}
\usepackage{color}

\usepackage{graphicx}
\usepackage[body={6.5in, 9in}, right=1in, top=1in]{geometry}
\usepackage{amssymb}
\usepackage{amsmath}
\usepackage[numbers,sort&compress]{natbib}

\numberwithin{equation}{section}
\newcommand{\paren}[1]{\left( #1 \right)}

\newcommand\ignore[1]{}
\def\one{{\,\hbox{1\kern-.8mm l}}}

\def\({\left(}
\def\){\right)}

\newcommand{\rmd}{\mathrm{d}}

\newcommand{\Cset}{{\,\,{{{^{_{\pmb{\mid}}}}\kern-.45em{\mathrm C}}}}}

\newcommand{\be}{\begin{equation}}
\newcommand{\bea}{\begin{eqnarray}}

\newcommand{\ee}{\end{equation}}
\newcommand{\eea}{\end{eqnarray}}

\newcommand{\ret}{\nonumber \\}

\def\mpl{M_{\rm Pl}}

\newcommand{\Comment}[1]{{}}
\definecolor{MyDarkBlue}{rgb}{0.15,0.15,0.45}
\usepackage[linktocpage=true]{hyperref}
\hypersetup{
colorlinks=true,
citecolor=MyDarkBlue,
linkcolor=MyDarkBlue,
urlcolor=MyDarkBlue,
}

\begin{document}
\def\thefootnote{\fnsymbol{footnote}}

\begin{center}
\Large{\textbf{The Physical Squeezed Limit: \\Consistency Relations at Order $q^2$}} \\[0.5cm]

\large{Paolo Creminelli$^{\rm a}$, Ashley Perko$^{\rm b}$, Leonardo Senatore$^{\rm b,c,d}$,\\ Marko Simonovi\'c$^{\rm e,f}$, and Gabriele Trevisan$^{\rm e,f}$}
\\[0.5cm]

\small{
\textit{$^{\rm a}$ Abdus Salam International Centre for Theoretical Physics\\ Strada Costiera 11, 34151, Trieste, Italy}}

\vspace{.2cm}

\small{
\textit{$^{\rm b}$ Stanford Institute for Theoretical Physics, Stanford University, Stanford, CA 94306}}

\vspace{.2cm}

\small{
\textit{$^{\rm c}$ Kavli Institute for Particle Astrophysics and Cosmology,\\ Stanford University and SLAC, Menlo Park, CA 94025}}

\vspace{.2cm}

\small{
\textit{$^{\rm d}$ CERN, Theory Division, 1211 Geneva 23, Switzerland}}

\vspace{.2cm}

\small{
\textit{$^{\rm e}$ SISSA, via Bonomea 265, 34136, Trieste, Italy}}
\vspace{.2cm}

\small{
\textit{$^{\rm f}$ Istituto Nazionale di Fisica Nucleare, Sezione di Trieste, 34136, Trieste, Italy}}

\end{center}

\vspace{.8cm}

\hrule \vspace{0.3cm}
\noindent \small{\textbf{Abstract}\\ In single-field models of inflation the effect of a long mode with momentum $q$ reduces to a diffeomorphism at zeroth and first order in $q$. This gives the well-known consistency relations for the $n$-point functions. At order $q^2$ the long mode has a physical effect on the short ones, since it induces curvature, and we expect that this effect is the same as being in a curved FRW universe. In this paper we verify this intuition in various examples of the three-point function, whose behaviour at order $q^2$ can be written in terms of the power spectrum in a curved universe. This gives a simple alternative understanding of the level of non-Gaussianity in single-field models. Non-Gaussianity is always parametrically enhanced when modes freeze at a physical scale $k_{\rm ph,\,f}$ shorter than~$H$:~$f_{\rm NL} \sim (k_{\rm ph,\,f}/H)^2$.}
\vspace{0.3cm}
\noindent
\hrule
\def\thefootnote{\arabic{footnote}}
\setcounter{footnote}{0}


\section{Introduction and main results}
In single-field inflation, correlation functions of the curvature perturbation $\zeta$ and of tensor modes satisfy general, model-independent relations in the limit in which one of the momenta (or the sum of some of the momenta) becomes very small compared to the others: the so-called consistency relations \cite{Maldacena:2002vr,Creminelli:2004yq,Cheung:2007sv,Creminelli:2012ed, Hinterbichler:2012nm,Senatore:2012wy,Hinterbichler:2013dpa,Goldberger:2013rsa,Chen:2013aj}. Physically, they are based on the observation that the effect of a long mode on the dynamics of the short ones reduces to a diffeomorphism. At zeroth and first order in gradients, the long mode can be removed by a suitable change of coordinates, so that the correlation function boils down to the effect of this diffeomorphism on the short modes. Single-field consistency relations are therefore completely model-independent and their violation would represent a clear smoking gun of multi-field models. On the other hand these relations are somewhat trivial, as they simply state that the long mode does not affect the short ones in a coordinate-independent way, and so they do not contain any dynamical information. Of course this does not hold at second order in gradients, because at this order the long mode induces curvature, which obviously cannot be erased by a change of coordinates.

In this paper we study single-field correlation functions in the limit in which one of the momenta, $q$, becomes soft, focusing on the leading ``physical'' effect, i.e.~at order $q^2$. At this order the long mode induces curvature and we expect that, after we take a proper average over directions, its effect is the same as being in a curved Friedmann-Robertson-Walker (FRW) universe. (We will verify this equivalence in section \ref{subsec.curvedfrw}.)
For the three-point function we will get
\be
\label{eq:schematic1}
\langle \zeta_{\vec q} \zeta_{\vec k_1} \zeta_{\vec k_2} \rangle_{q \to 0,\,{\rm avg}}{=}  P(q)\;\frac{2}{3} \cdot q^2 \cdot \frac{\partial}{\partial \kappa} \langle \zeta_{\vec k_1} \zeta_{-\vec k_1} \rangle \;,
\ee
where $P$ is the power spectrum, $\kappa$ is the spatial curvature of the FRW and the subscript ${\rm avg}$ indicates that an angular average has been taken.
We will verify this relation in various examples. Notice that the equation above is rather different from a standard consistency relation, which connects observable correlation functions in our universe, so that one can check whether it is experimentally violated. Here the right-hand side of Eq.~\eqref{eq:schematic1} contains the power spectrum in a {\em curved} universe, which   
cannot be measured independently. Therefore one cannot check experimentally whether the relation holds or not. Its interest is mostly conceptual since it allows to understand in simple physical terms the origin of non-Gaussianities in the regime of squeezing, ${\cal O}(q^2)$, where they are potentially large. 
Indeed the equivalence between a long mode and a curved universe allows to deduce some general property of correlation functions.  

\begin{figure}
\centering
\includegraphics[width=5.6in]{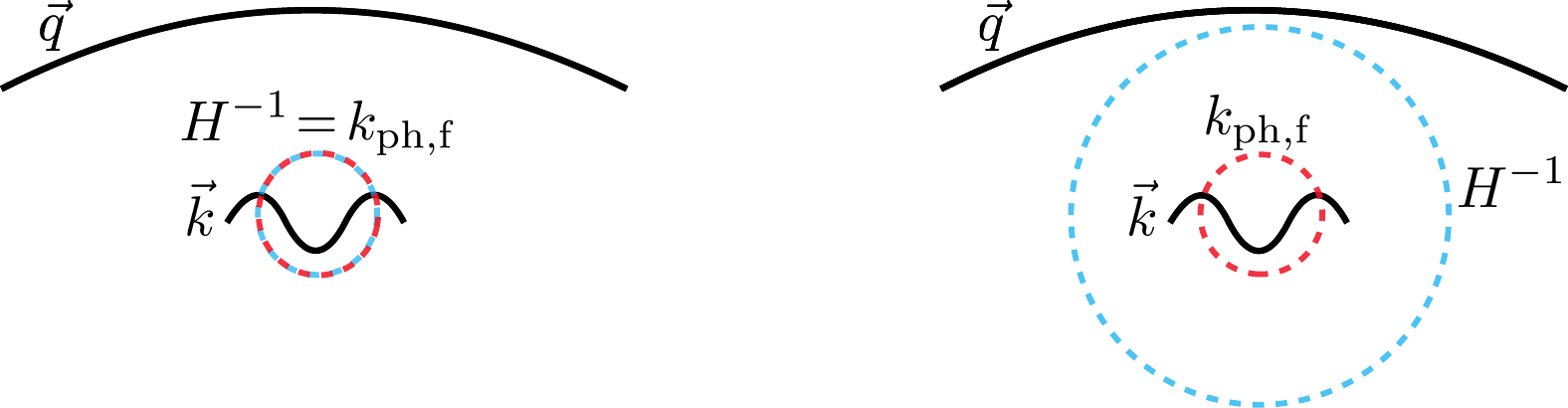}
\caption{\small \label{fig:freezing}Parametric dependence of non-Gaussianity with the freezing scale. On the left the freezing scale is the Hubble scale. On the right the freezing scale is parametrically shorter than Hubble: in this case the effect of the curvature induced by the long mode is bigger, since it has to be compared with $H^2$ in the Friedmann equation.}
\end{figure}

A general conclusion one can draw from \eqref{eq:schematic1} is the following. {\em The level of non-Gaussianity is always parametrically enhanced when modes freeze at a wavelength shorter than the Hubble scale}. More precisely
\be
\label{eq:parametric}
f_{\rm NL} \sim \left(\frac{k_{\rm ph,\,f}}{H}\right)^2 \;,
\ee
where $k_{\rm ph,\,f}$ is the physical freezing scale. 

The argument for Eq. (\ref{eq:parametric}) is very simple and it is illustrated in Figure~\ref{fig:freezing}. Let us consider a squeezed triangle with a given ratio between the long and the short modes $\varepsilon \equiv \frac{q}{k} \ll 1$. The power spectrum of short modes is imprinted at freezing and at this moment the long mode has momentum $\varepsilon \,k_{\rm ph,\,f}$. The effect of the long mode is the same as being in a curved FRW, with curvature $\kappa \sim {q^2}/{a^2}\cdot \zeta_q \sim \varepsilon^2 {k^2}/{a^2} \cdot \zeta_q$. Curvature will modify the spectrum of the short modes through its effect in the Friedmann equation and its consequent change of the inflaton speed $\dot\phi$. The relative correction introduced in the Friedmann equation, and thus in the power spectrum for the short modes, is obtained by comparing the spatial curvature with $H^2$ at freezing time: $({q^2}/{a_f^2}\cdot {\zeta_q})/{H^2} \simeq \varepsilon^2 k_{\rm ph,\,f}^2 {\zeta_q}/{H^2}$. This implies that the three-point function behaves as  $(k_{\rm ph,\,f}/H)^2 (q/k)^2 P(q) P(k)$ in the squeezed limit, i.e.~$f_{\rm NL} \sim (k_{\rm ph,\,f}/H)^2$. Although the derivation works only in the squeezed limit, we expect by continuity (since at this order we are capturing the physical effect of the long mode) that it gives the correct parametric dependence also when modes become comparable\footnote{Strictly speaking, $f_{\rm NL}$ is defined as proportional to the value of the 3-point function in the equilateral configuration. In this regime, the derivative expansion in $q/k$ clearly does not apply and we therefore have nothing to say for this configuration. Here for $f_{\rm NL}$ we simply mean a typical value of the 3-point function apart from points where it accidentally cancels, or equivalently the observational limit on $f_{\rm NL}$ that would be obtained by a dedicated analysis of the 3-point function in the data. Indeed, since at order $q^2$ we are capturing a true physical effect, it is hard to imagine that there are cancellations for all configurations so that the size of the non-Gaussianities is parametrically different than the one obtained by extrapolating our expression (\ref{eq:parametric}) to the equilateral limit. This expectation is indeed confirmed by explicit calculations of the 3-point function in all triangular configurations. }.

The relation \eqref{eq:parametric} obviously works for models with reduced speed of sound \cite{Garriga:1999vw,Chen:2006nt,Cheung:2007st,Alishahiha:2004eh}: 
\be
{\rm Reduced\;} c_s: \qquad k_{\rm ph,\,f} = \frac{H}{c_s}  \quad \Rightarrow \quad f_{\rm NL} \sim \left(\frac{k_{\rm ph,\,f}}{H}\right)^2 \sim \frac{1}{c_s^2} \;.
\ee
Indeed we are going to check Eq.~\eqref{eq:schematic1} both at order $1/c_s^2$, when one can neglect the curved spatial geometry and take only into account its effect on the scale factor and the inflaton speed (see section \ref{subsec.eftsmallcs}), and at order one, when the check is more complicated since one has to relate the three-point function calculation done in terms of Fourier modes with the spectrum in the curved geometry (see section \ref{subsec.eftfull}). 
The same argument works when dissipative effects are present. In this class of  models \cite{LopezNacir:2011kk} a parameter $\gamma \gg H$ plays the role of an effective friction and the freezing happens when $c_s^2 k^2/a^2 \sim \gamma H$.
This means that Eq.~\eqref{eq:parametric} reads
\be
{\rm Dissipative:} \qquad k_{\rm ph,\,f} = \frac{\sqrt{\gamma H}}{c_s} \quad \Rightarrow \quad f_{\rm NL} \sim \left(\frac{k_{\rm ph,\,f}}{H}\right)^2 \sim \frac{\gamma}{H c_s^2} \;.
\ee
This matches the explicit calculations of \cite{LopezNacir:2011kk}.
The argument leading to Eq.~\eqref{eq:parametric} does not rely on a linear dispersion relation for the inflaton perturbations and indeed it gives the correct estimate for Ghost Inflation \cite{ArkaniHamed:2003uz} where the dispersion relation is of the form $\omega = k^2/M$. This gives
\be
{\rm Ghost \;Inflation:} \qquad k_{\rm ph,\,f} = (H M)^{1/2}  \quad \Rightarrow \quad f_{\rm NL} \sim \left(\frac{k_{\rm ph,\,f}}{H}\right)^2 \sim \frac{M}{H} \;.
\ee
The check of Eq.~\eqref{eq:schematic1} for Ghost Inflation is deferred to appendix \ref{sec:ghost}. A peculiar example is the one of Khronon Inflation \cite{Creminelli:2012xb}: in this case the power spectrum is not affected by the background curvature at order $1/c_s^2$ since in the quadratic action the scale factor does not appear at this order and the speed of the inflaton background is immaterial due to the field redefinition symmetry. Indeed, we check in appendix \ref{sec:khronon} that the $1/c_s^2$ terms of the three-point function at order $q^2$ vanish, when the proper angular average is taken.  
The general argument above implies that one has completely non-Gaussian perturbations, i.e.~strong coupling, when freezing occurs at a sufficiently short scale
\be
{\rm Strong\;coupling:} \qquad \left(\frac{k_{\rm ph,\,f}}{H}\right)^2 \sim P_\zeta^{-1/2} \;.
\ee

It is worth noticing that our arguments cannot be extended beyond the $q^2$ order. At order $q^3$ one {\em cannot} interpret the three-point function as a calculation of the two-point function on a long classical background. Indeed, at this order one cannot neglect the time-dependent phase of the long mode and therefore one cannot describe the long mode classically. Moreover, one cannot neglect anymore the probability that two short modes combine to give a long one. 

Some of these arguments can be straightforwardly extended to the trispectrum and higher correlation functions. For example one can apply twice the argument leading to Eq.~\eqref{eq:schematic1} to conclude that the amplitude of the four-point function will behave as
\be
\tau_{\rm NL} \sim \left(\frac{k_{\rm ph,\,f}}{H}\right)^4 \;.
\ee
This corresponds to $1/c_s^4$ for models with reduced speed of sound and $(M/H)^2$ for Ghost Inflation. 

Before getting into details, it is important to stress a drawback of the point of view we take in this paper. Here we are associating non-Gaussianities with the effect of the curvature, by setting the long mode fluctuation of the inflation to zero with a choice of coordinates. In this way, the interactions between the long mode and the short modes appear of gravitational origin.  We know that the three-point function can also be obtained (up to slow-roll corrections) in the decoupling limit. In particular in models with $c_s \ll 1$ freezing happens at scales much shorter than the curvature and large non-Gaussianities can be understood in flat space. This simplification is somewhat obscured in our language.

\section{From $\zeta$-gauge to a curved FRW universe} \label{subsec.curvedfrw}
The consistency relations at zeroth and first order in the Taylor expansion in $q$ of $\zeta_L$, the long wavelength component of $\zeta$, have already been exhaustively considered in \cite{Maldacena:2002vr,Creminelli:2004yq,Cheung:2007sv,Creminelli:2012ed, Hinterbichler:2012nm,Senatore:2012wy,Hinterbichler:2013dpa,Goldberger:2013rsa}. Thus in this paper we will neglect constant and gradient terms, and consider only the Taylor expansion at second order,
\be
\zeta_L(\vec x,t)= \zeta_{L}|_0(t)+\frac{1}{2} \partial_i \partial_j \zeta_L|_0 \,x^i x^j \ .
\label{taylor}
\ee
The time dependence of $\zeta_L$ is fixed by the equation of motion of $\zeta$. Focusing on models with reduced speed of sound we have
\be
\label{eq:EOM}	
\partial_t \left( \epsilon a_0^3   \dot \zeta_L\right)-\epsilon a_0 c_s^2  \,\partial^2 \zeta_L=0 \ ,
\ee
where $a_0$ is the scale factor of the flat FRW universe and $\epsilon=-\dot H/H^2$ is the slow-roll parameter. Notice that we cannot neglect the time dependence of $\zeta_L$ since it is of order $q^2$, if not larger. In this paper we will assume that the time dependence of $\zeta$ starts at order $q^2$. Indeed the differential equation above admits, in the limit $q \to 0$, also a time-dependent solution for~$\zeta$. This solution is relevant when the background solution has not yet reached the attractor. This leads to a violation of the standard Maldacena consistency condition~\cite{Namjoo:2012aa}, though we expect that generalized consistency conditions hold even for these cases by considering that the universe in the presence of a long mode can now be locally described as a flat anisotropic one~\cite{Senatore:2012wy,Pimentel:2012tw}. In this paper we focus only on the case in which the background has already reached the attractor and the modes are in the Bunch-Davies vacuum, so that the time-dependence of $\zeta$ outside the horizon is of order $q^2$. Substituting in Eq.~(\ref{taylor}) and solving for $\zeta_{L}(t)$ at first order in the slow-roll parameters, we find that the long mode at quadratic order is (\footnote{In solving the differential equation \eqref{eq:EOM} the lower bounds of the integrals give contributions proportional to $q^2$ that are either constant in time or decaying as $1/a^3$. The first term can always be reabsorbed in the definition of the asymptotic value of $\zeta$, while we set the other term to zero since it would correspond to a decaying mode proportional to $q^2$, which is absent for any time-translation invariant state, such as the vacuum state.})
\be 
\zeta_L(\vec x,t)=-\frac{c_s^2 \partial^2\zeta_L|_0}{2 a_0^2 H_0^2}(1-2\epsilon)+\frac{1}{2} \partial_i \partial_j \zeta_L|_0 \,x^i x^j \,.
\label{long}
\ee

In order to show that at this order the long mode is equivalent to being in a spatially curved FRW, it is enough to consider the perturbed metric in $\zeta$-gauge at linear order (see for example \cite{Chen:2006nt})
\begin{equation}	
\begin{split}
g_{00}&=-1-2\frac{\dot \zeta_L}{H_0}  \ ,\\
g_{0i}&=-\frac{\partial_i \zeta_L }{H_0}+\frac{a_0^2 \epsilon}{c_s^2}\frac{\partial_i \dot \zeta_L}{\partial^2}  \ , \\
g_{ij}&=a_0^2 \paren{1+2\zeta_L} \delta_{ij} \ ,
\label{flat_metric}
\end{split}
\end{equation}
and make a gauge transformation to rewrite it in the usual FRW form (i.e.~by setting $g_{00}=-1$ and $g_{0i}=0$). Under the coordinate transformation $x^\mu\to \tilde x^\mu=x^\mu+\xi^\mu(\tilde x)$, the metric transforms as
\begin{equation}	
\begin{split}
\tilde g_{00}&= g_{00}+2\dot \xi^0 \ , \\
\tilde g_{0i}&= g_{0i} -a_0^2\dot \xi^i+\partial_i \xi^0 \ , \\
\tilde g_{ij}&= g_{ij}-a_0^2\partial_i \xi^j-a_0^2\partial_j \xi^i-2 a_0^2 H_0 \xi^0 \delta_{ij} \ ,
\end{split}
\end{equation}
and we can set $\tilde g_{00}=-1$ and $\tilde g_{0i}=0$ by choosing the transformation parameters to be
\begin{align}
\xi^0&= \int^t \rmd t' \frac{\dot \zeta_L}{H_0} = -\frac{c_s^2 \partial^2\zeta_L|_0}{2 a_0^2 H_0^3}\paren{1-\frac{3}{2}\epsilon}\ , \ret
\xi^i&=-\int^t \rmd t' \frac{\partial_i \zeta_L}{a_0^2 H_0}\paren{1-\epsilon}=\frac{\partial_i \partial_j \zeta_L|_0 \tilde x^j}{2 a_0^2 H_0^2} \ .
\label{xi}
\end{align}
The effect of the long mode is now entirely encoded in the spatial part of the metric 
\begin{align}
\tilde g_{ij}=a_0^2\paren{\delta_{ij}-\frac{\partial_i \partial_j \zeta_L|_0}{a_0^2H_0^2}+\epsilon \frac{c_s^2 \partial^2\zeta_L|_0}{2 a_0^2 H_0^2}\delta_{ij}}\paren{1+\partial_a \partial_b \zeta_L|_0 \tilde x^a \tilde x^b} \ ,
\label{curved_metric}
\end{align}
which, once averaged
over angles, takes the familiar form
\begin{equation}
 \tilde{ g}_{ij,\,{\rm avg}} {=} a^2\paren{1-\frac{1}{2}\kappa \tilde x^2} \ ,
 \label{iso_metric}
 \end{equation}
 where the spatial curvature $\kappa$ and the scale factor $a$ in the curved universe are given by
 \begin{align}
 \kappa&=-\frac{2}{3}\partial^2 \zeta_L|_0  \ , \ret
 a^2&=a_0^2 \paren{1+\frac{\kappa}{2 a_0^2 H_0^2} -\epsilon  \frac{3 c_s^2\kappa }{4 a_0^2 H_0^2}}\ .
  \label{iso_scale}
\end{align}
It is important to stress that the change of coordinates \eqref{xi} decays away at late time. This implies that, since we are interested in correlation functions at late times, one can forget about the change of coordinates and just treat the long mode as adding spatial curvature to the FRW solution.

Eventually the consistency condition at this order can be written as
\be
\label{eq:schematic}
\langle \zeta_{\vec q} \zeta_{\vec k_1} \zeta_{\vec k_2} \rangle'_{q \to 0,\,{\rm avg}}{=}  P_{\zeta}(q) \;\frac{2}{3}\cdot q^2 \cdot \frac{\partial}{\partial \kappa} \langle \zeta_{\vec k_1} \zeta_{-\vec k_1} \rangle'_\kappa \; ,
\ee 
where the subscript $\kappa$ means that the two-point function is calculated in the curved universe and primes on correlation functions indicate that a $(2\pi)^3 \delta(\sum_i \vec k_i)$ of momentum conservation has been removed. As stated above, the assumption used to derive this formula is the fact that the background solution is on the attractor, so that the time-dependence of the long $\zeta$ mode starts at order $q^2$. Additionally, in deriving Eq. (\ref{iso_scale}) we used the slow-roll approximation, though we expect the relationship to hold at all orders in the slow-roll expansion. Our consistency condition should indeed hold for all FRW backgrounds that are accelerating so that modes freeze after horizon crossing. This implies that the effect of the long mode on the short ones is captured by a derivative expansion of the long mode at the time when the short modes cross the horizon. This ceases to be the case for decelerating FRW backgrounds, when modes re-enter the Hubble scale, and so the derivative expansion is not applicable.

One can generalize these arguments without taking the angular average and consider a curved anisotropic (but homogeneous) universe. In this case Eq.~\eqref{eq:schematic} becomes
\be
\langle \zeta_{\vec{q}}\zeta_{\vec{k}_1}\zeta_{\vec{k}_2}\rangle_{q \to 0}' = P_{\zeta}(q) \  q_i q_j \frac{\partial }{\partial (q_i q_j\zeta_{\vec q})}  \langle \zeta_{\vec{k}}\zeta_{-\vec{k}} \rangle'_{\rm local} \ ,
\ee
where $\langle\dots \rangle_{\rm local}$ means that the two-point function is calculated in a locally anisotropic curved universe. We will study the anisotropic case in appendix~\ref{app:gauge}.

\section{Checks of the consistency relation}

In this section we are going to explicitly check our consistency relation in models with reduced speed of sound. We are going to work in the framework of the Effective Field Theory of Inflation (EFTI) developed in \cite{Cheung:2007st}, while in appendix \ref{app:k-infl} we derive some of the results using a k-inflation action \cite{Garriga:1999vw}. The EFTI is built on the fact that during inflation there is a physical clock that breaks spacetime diffeomorphism invariance while preserving time-dependent spatial diffeomorphisms. Thus the dynamics of the inflaton is captured in an effective action derived by writing all terms consistent with the residual symmetries around a given cosmological background. The most general effective action around the flat FRW background at lowest order in derivatives and up to cubic order in perturbations is~\cite{Cheung:2007st}
\begin{align}
S&=\int \rmd^4 x \sqrt{-g}  \mpl^2\left[-\paren{3H_0^2+\dot H_0}+\dot H_0 g^{00} \right. \ret
&\left. \qquad+\frac{1}{2}\paren{\frac{c_s^2-1}{2 c_s^2}}\dot H_0 \paren{g^{00}+1}^2+\frac{1}{6}\paren{\frac{1-c_s^2}{2c_s^4}c_3}\dot H_0 \paren{g^{00}+1}^3+\ldots \right]\ , 
\label{flat_action}
\end{align}
where $a_0$ is the unperturbed history for the spatially flat FRW, $c_s$ is the speed of sound of the perturbations, and $c_3$ is a dimensionless parameter not fixed by symmetries. For simplicity we take $c_s$ and $c_3$ to be time independent. To describe the fluctuations around a flat FRW background it is useful to introduce the Goldstone boson $\pi$ that nonlinearly realizes time diffeomorphisms. Following~\cite{Cheung:2007st}, at leading order in slow-roll, we can neglect metric fluctuations and go to the decoupling limit. The cubic action for $\pi$ becomes
\begin{align}\label{Eq.cub_int}
S^{(3)} = \int \rmd^4x \;a_0^3 \left(-\frac{\dot H_0 \mpl^2}{c_s^2}\right) \left[ - (1-c_s^2) \dot\pi  \frac{(\partial_i\pi)^2}{a_0^2} + (1-c_s^2) \left( 1+ \frac{2}{3}\frac{c_3}{c_s^2} \right) \dot\pi^3 \right] \;.
\end{align}
Now it is straightforward to calculate the three-point function. Using $\zeta=-H_0\pi$, we get the following expression 
\begin{align}
\langle \zeta_{\vec k_1} \zeta_{\vec k_2} \zeta_{\vec k_3} \rangle' = & \left( \frac{H_0^2}{4\epsilon c_s M_{pl}^2} \right)^2 \frac{1}{\prod_i 2k_i^3} \left[ \left(1-\frac{1}{c_s^2}\right) \left(c_3+ \frac 32 c_s^2\right) \frac{32k_1^2k_2^2k_3^2}{k_t^2} \right. \nonumber \\
& \left. - 4 \left(1-\frac{1}{c_s^2}\right) \left(12 \frac{k_1^2k_2^2k_3^2}{k_t^3} - \frac{8}{k_t}\sum_{i>j} k_i^2 k_j^2 + \frac{4}{k_t^2}\sum_{i\neq j} k_i^2 k_j^3 + \sum_i k_i^3 \right) \right] \;,
\end{align}
with $k_t \equiv k_1 + k_2 + k_3$.
In order to find the squeezed limit we can expand the expression above using 
\be
\label{symexp}
\vec k_1 = \vec k - \frac 12 \vec q\;, \quad \vec k_2 = -\vec k - \frac 12 \vec q \;, \quad \vec k_3 = \vec q\;,\quad q\ll k\;,
\ee
because in this way it is evident that the three-point function does not have corrections that are linear in $q$. Expanding up to second order in $q$ we find
\be
\label{squeezed3pf}
\langle \zeta_{\vec q} \zeta_{\vec k_1} \zeta_{\vec k_2} \rangle'_{q\to 0} = P_\zeta(q) P_\zeta(k) \left( 1- \frac{1}{c_s^2} \right) \left[ \left( 2+ \frac 12 c_3 + \frac 34 c_s^2 \right)\frac{q^2}{k^2} - \frac 54 \frac{(\vec k \cdot \vec q)^2}{k^4}\right] + \mathcal O\left(\frac{q^3}{k^3} \right) \;.
\ee

What is left to do in order to verify the consistency condition (\ref{eq:schematic}) is to calculate the two-point function in the spatially curved FRW. What makes this calculation nontrivial is that the two-point function in the curved universe must be expressed in terms of flat universe parameters (the speed of sound $c_s$ and $c_3$) because we want to compare it to the three-point function. In the rest of this section we are going to show how this can be done. First, we will derive the quadratic action for short modes in a curved FRW and match the parameters with the flat ones. Then, we will calculate the two-point function and show that the consistency relation is satisfied.

\subsection{Quadratic action for short modes in a curved FRW universe} \label{subsec:eftaction}
In order to calculate the two-point function in a curved FRW we need the action for the short modes. The most straightforward thing to do is to write the EFTI for a curved FRW~\cite{Cheung:2007st},
\begin{align}
\tilde S &=\int \rmd^4\tilde{x}  \sqrt{-\tilde g} \ \mpl^2 \left[-\paren{3 H^2+\dot { H}+2\frac{\kappa}{a^2}}+\paren{\dot { H}-\frac{\kappa}{a^2} }\tilde g^{00} +\frac{1}{2}\paren{\frac{\tilde c_s^2-1}{2 \tilde c_s^2}}\paren{\dot { H}-\frac{\kappa}{a^2} }\paren{\tilde g^{00}+1}^2 \right] \;,
\label{curved_eft}
\end{align}
where tildes indicate curved quantities, and introduce the Goldstone $\tilde \pi_S$ via the Stueckelberg trick $\tilde t \to \tilde t +\tilde \pi_S$. This procedure leads to the following action
\begin{equation}
\label{curved_pi}
\begin{split}
\tilde S_{\rm avg}{=}-\int \rmd^4\tilde x \     \sqrt{-\tilde g} \paren{\dot { H}-\frac{\kappa}{a^2} } \mpl^2&\left[\frac{1}{\tilde c_s^2}\dot{\tilde \pi}_S^2 -\partial^i \tilde \pi_S\partial_i \tilde \pi_S+3\dot H\tilde \pi_S^2\right] \ .
\end{split}
\end{equation}

Note however that  $\tilde c_s$, which represents the speed of propagation of the $\tilde\pi_S$ waves, is a free parameter which is not determined by the background cosmology and has to be related to $c_s$ and $c_3$, the parameters of the effective action in the flat universe. The relation can be found by noticing that the speed of propagation is a physical quantity that can be measured on short scales, where the universe is locally flat. Thus the speed of propagation must be the same in the curved and in the flat EFTI, provided that in the latter we take into account the background of the long wave $\pi_L$. The speed of $\pi_S$ waves over the background of the long mode can be read directly from the action in the flat universe Eq.~(\ref{flat_action}) once in the interaction Lagrangian in Eq.~(\ref{Eq.cub_int}) we replace $\pi=\pi_L+\pi_S$. We can go on short scales and neglect the space dependence of $\pi_L$. The relevant action reads
\begin{align}
S = \int \rmd^4x \ \epsilon \mpl^2 H_0^2 a_0^3 \frac{1}{c_s^2} \left[\left( 1+ (1-c_s^2) \left( 1+ \frac{2}{3}\frac{c_3}{c_s^2} \right)3 \dot\pi_L \right)\dot\pi^2_S - c_s^2 \left(1 + \frac1{c_s^2}(1-c_s^2) \dot{\pi}_L \right) \frac{(\partial_i\pi_S)^2}{a_0^2} \right] \;.
\end{align}
If we consider the long mode of $\pi$: $\pi_L=-\zeta_L/H_0$, where $\zeta_L$ is given in Eq.~(\ref{long}), we obtain the following quadratic action for $\pi_S$
\begin{align}
S^{(2)} = \int \rmd^4x \ \epsilon \mpl^2 H_0^2 a_0^3 \frac{1}{c_s^2}\left[ \left( 1+(1-c_s^2) \left( 3c_s^2+ 2 c_3 \right) \frac{3 \kappa}{2a_0^2H_0^2}\right)\dot\pi^2_S - c_s^2\left(1+(1-c_s^2) \frac{3 \kappa}{2a_0^2H_0^2} \right) \frac{(\partial_i\pi_S)^2}{a_0^2}   \right] \;.
\end{align}
From this we can read the speed of sound in the flat EFTI and this must be the same as $\tilde c_s^2$:
\begin{equation}\label{Eq.curved_cs}	
\begin{split}
\tilde c_s^2=c_s^2 \frac{ \left(1+(1-c_s^2) \frac{3 \kappa}{2a_0^2H_0^2} \right)}{ \left( 1+(1-c_s^2) \left( 3c_s^2+ 2 c_3 \right) \frac{3 \kappa}{2a_0^2H_0^2}\right)}\simeq c_s^2\left( 1+\frac{3\kappa}{2a_0^2H_0^2}\left( c_s^2-1\right)\left( 3c_s^2+2c_3-1\right)\right) \ ,
\end{split}
\end{equation}
where in the second passage we have expanded up to linear order in $\kappa$, as we always do in this paper.
This fixes the speed of sound in the curved EFTI. 

We can now plug this value for $\tilde c_s$ and use the scale factor of Eq. (\ref{iso_scale}) in the action of Eq. (\ref{curved_pi}) to obtain the action for the short modes. At first order in $\kappa$ and $\epsilon$ it reads
\begin{align}\label{Eq.curved_action}
 S_{\rm avg}{=}\int \rmd^4 x \    \epsilon \mpl^2 H_0^2 a_0^2   &  \paren{1-\frac{3}{4}\kappa x^2} \left[\frac{1}{c_s^2} \paren{1+ \frac{3\kappa}{4 a_0^2 H_0^2} \paren{1+4c_3+10c_s^2-4c_3c_s^2- 6c_s^4}} { {\pi}}_S'^2 \right. \ret
& \quad \left. -\paren{1+\frac{1}{2}\kappa x^2}\left(1+\frac{\kappa}{4a_0^2H_0^2}\paren{7+6 c_s^2} \right)(\partial_i\pi_S)^2 + 3\kappa  {\pi}^2_S\right] \ ,
\end{align}
where we have used conformal time and suppressed tildes for simplicity. Starting from this action, in the next sections, we will calculate the two-point function of the short modes and check the consistency relation.

However, before doing so, it is important to notice that the two Goldstone modes -- $\tilde \pi_S$ in the curved universe and $\pi_S$ in the flat universe -- are different given that we are expanding around different backgrounds. To relate the two it is sufficient to consider the change of coordinates in Eq.~(\ref{xi}). Since $\tilde \pi_S$ is the perturbation around $\tilde t$ and $\pi_S$ is the perturbation around $t$, the relation between the two is 
\begin{equation}	
\begin{split}
\tilde \pi_S=(1+\dot \xi^0)\pi_S \ ,
\end{split}
\end{equation}
where the time dependence of $\xi^0$ is $a_0^{-2}$. Eventually the effect of curvature and anisotropy goes to zero at late times and thus $\tilde \pi_S= \pi_S$ when we evaluate correlation functions. This also means that we can directly relate correlation functions of $\tilde \pi_S$ in the local curved universe to correlation functions of global quantities. The nonlinear relationship between $\zeta_S$ and $\pi_S$ in the flat FRW universe is
\be
\zeta_S=-H_0 \pi_S+\frac{1}{2}\dot H_0 \pi_S^2 \ ,
\label{zeta_pi_global}
\ee
at second order, plus terms with derivatives of $\pi_S$ that vanish after horizon crossing~\cite{Cheung:2007sv}. The $\pi_S^2$ term in Eq.~(\ref{zeta_pi_global}) is slow-roll suppressed so we can neglect its contribution to the three-point function. We then find that at late times the local $\tilde \pi_S$ is related to the global $\zeta_S$ as
\be
\zeta_S=-H_0 \tilde{\pi}_S \ ,
\label{zeta_pi}
\ee
to leading order in the slow-roll parameter $\epsilon$.

\subsection{The two-point function at leading order in $1/c_s^2$} \label{subsec.eftsmallcs}

Now that we have the action for the short modes in the curved universe (\ref{Eq.curved_action}), we can proceed with the calculation of the two-point function and check the consistency relation. In this subsection we will focus on the case of leading order in $1/c_s^2$ for an isotropic long mode. When the speed of sound is parametrically smaller than $1$ the relevant effects of the long mode are the modifications of the scale factor, the Hubble constant and the speed of sound. As we discussed in the introduction, these affect the 2-point function of the short $k$-modes proportionally to $\kappa/H^2\sim ({q^2}/{a_f^2}\cdot {\zeta_q})/{H^2} \sim \varepsilon^2 k_{\rm ph,\,f}^2 {\zeta_q}/H^2\sim\varepsilon^2 \zeta_q/c_s^2$, where $q$ is the long mode and $\varepsilon=q/k\ll1$ is the squeezing parameter. On the contrary, geometrical effects scale as $\kappa\, x^2\,\sim  q^2/k^2\cdot \zeta_q\sim \varepsilon^2 \zeta_q$, and are therefore not enhanced for $c_s\ll1$.

Given that at order $1/c_s^2$ geometrical effects and the mass term do not contribute, the leading part of the action \eqref{Eq.curved_action} is given by
\begin{align}
 S_{\rm avg}&{=}\int \rmd^4 x \    \epsilon   \mpl^2 H_0^2  a_0^2\left[\frac{1}{c_s^2} \paren{1+ \frac{3 \kappa}{4a_0^2 H_0^2}(4c_3+1)}  {\pi '}^2  -\paren{1+\frac{7\kappa}{4a_0^2H_0^2}} (\partial_i \pi)^2\right] \ .
\end{align}
In this approximation space can be considered to be flat and one can get the power spectrum for the short modes using the standard quantization of the scalar field. After going to Fourier space the equation of motion for the short modes is
\be
\pi ''_{\vec k} - \frac{2}{\eta}\pi '_{\vec k} + c_s^2k^2 \pi_{\vec k} + \frac{\kappa}{4a_0^2  H_0^2} \left( 3(4c_3+1){\pi ''_{\vec k}} + 7 c_s^2 k^2 \pi_{\vec k} \right)= 0 \;,
\ee
and it can be easily solved perturbatively in $\kappa$. The zeroth order solution is proportional to the standard de Sitter modes, but the normalization can be in principle different. In order to fix it we have to demand that the commutator of the field $\pi$ and the generalized momentum $\Pi$ is given by
\be
[\pi_{\vec k}(\eta),\Pi_{\vec k'} (\eta)] = i \delta(\vec k + \vec k')\;.
\ee
After fixing the normalization it is easy to find the two-point function in the late time limit. The result is
\be
\langle \zeta_{\vec k} \zeta_{-\vec k} \rangle'_\kappa = P_\zeta(k)\left( 1- \frac{19+6c_3}{8c_s^2k^2}\kappa \right) \;.
\ee
We can use this result to get the contribution to the squeezed limit of the three-point function. We only have to average over the long mode and recall that $\kappa=\frac 23 q^2\zeta_{\vec q}$
\be
\langle \zeta_{\vec q} \zeta_{\vec k_1} \zeta_{\vec k_2} \rangle'_{q\to 0} = - P_\zeta(q) P_\zeta(k) \frac{1}{c_s^2} \frac{19+6c_3}{12} \frac{q^2}{k^2}\;.
\ee
Comparing this result with \eqref{squeezed3pf}, we can see that two expressions are the same at the leading order in $1/c_s^2$ expansion if the proper angular average is taken into account. 

\subsection{The two-point function at all orders in $c_s$} \label{subsec.eftfull}
Similar arguments can be used to recover the squeezed limit of the three-point function at all orders in $c_s$. In this case we have to use the full action \eqref{Eq.curved_action}. The effects that induce the change of the two-point function are coming from the mass term, the curvature of the spatial slices (geometrical effects) and the change of the scale factor, the Hubble constant and the speed of sound induced by the long mode. Given that everything is perturbative in $\zeta_L$ we will treat these two effects separately.

Let us for the moment focus on the mass term, the change of $a$, $H$ and $c_s$ and neglect geometrical effects.  
This simplifies the calculation since we can still use Fourier transform and standard quantization procedure as before to obtain the power spectrum. 
The equation of motion obtained from \eqref{Eq.curved_action} neglecting the geometrical effects can be solved perturbatively in $\kappa$ and the normalization can be fixed in the same way as in the previous subsection. Looking at the late time limit of the solution, the two-point function we get is
\be
\langle \zeta_{\vec k} \zeta_{-\vec k} \rangle'_\kappa = P_\zeta(k)\left[ 1+\frac{3}{2} \left( \frac{1}{2}+\frac{c_3}{2} - \frac{19}{12c_s^2} - \frac{c_3}{2c_s^2} + \frac{3}{4}c_s^2 \right) \frac{\kappa}{k^2}\right] \;.
\ee
The squeezed limit of the three-point function can be obtained straightforwardly by correlating with the long mode, which gives
\be
\label{eq:nogeo}
\langle \zeta_{\vec q} \zeta_{\vec k_1} \zeta_{\vec k_2} \rangle'_{q\to 0} = P_\zeta(q) P_\zeta(k) \left( \frac{1}{2}+\frac{c_3}{2} - \frac{19}{12c_s^2} - \frac{c_3}{2c_s^2} + \frac{3}{4}c_s^2 \right) \frac{q^2}{k^2} \;.
\ee

Finally, we can focus on the contribution to the two-point function coming from the geometrical effects. We can read from \eqref{Eq.curved_action} that the relevant part of the action is
\begin{align}
S &=\int \rmd^4 x \    \epsilon \mpl^2 H_0^2 a_0^2     \paren{1-\frac{3}{4}\kappa x^2} \left[\frac{1}{c_s^2} {\pi '}^2 -\paren{1+\frac{1}{2}\kappa x^2} (\partial_i \pi)^2 \right] \ .
\end{align}
In principle, we can proceed as before solving the equation of motion. However, given that we are not in flat space anymore, we cannot expand $\pi$ in Fourier modes. The easiest way to calculate the two-point function is to treat the terms proportional to the curvature as an interaction. Then we can use the standard in-in calculation which guarantees the correct choice of the normalization and the choice of the vacuum for the modes. The interacting Hamiltonian is given by
\be
H_{int} =\int \rmd^4 x \    \epsilon  \mpl^2 H_0^2 a_0^2     \left[ \frac{3}{4c_s^2}\kappa x^2{\pi '}^2 - \frac{1}{4}\kappa x^2  (\partial_i \pi)^2  \right] \;,
\ee
and the contribution to the two-point function is
\be
\delta \langle \pi(x_1) \pi(x_2) \rangle_\kappa = -2\mathrm{Re} \left[ i \int^\eta \rmd^3\vec x \rmd \eta'\;\pi(x_1) \pi(x_2) H_{int}(\vec x,\eta') \right] \;, 
\ee
where the contour of integration has been implicitly rotated to pick up the interacting vacuum in the past.
At the end we want to write the result in momentum space. Using $x^2=\int \frac{\rmd^3 \vec p}{(2\pi)^3}(-\nabla^2_p \delta(\vec p))e^{i\vec p\cdot \vec x}$, we get the following equality
\begin{align}
\label{2pfreal}
& \int \rmd^3\vec k_1 \rmd^3\vec k_2 \langle \pi_{\vec k_1} \pi_{\vec k_2} \rangle e^{i\vec k_1\cdot\vec x_1 + i\vec k_2\cdot \vec x_2} = \int \rmd^3\vec k_1 \rmd^3\vec k_2 \; F(\vec k_1, \vec k_2) \nabla_{k_{1}}^2\delta(\vec k_1+ \vec k_2) e^{i\vec k_1\cdot\vec x_1 + i\vec k_2\cdot \vec x_2}  \;,
\end{align}
where the function $F(\vec k_1,\vec k_2)$ is
\be
F(\vec k_1, \vec k_2) = (2\pi)^3\; \mathrm{Re} \left[ i \epsilon\kappa \mpl^2 \; \pi_{\vec k_1}^* \pi_{\vec k_2}^* \int^\eta \frac{\rmd \eta'}{\eta'^2}  \left( \frac{3}{c_s^2}\pi'_{\vec k_1}\pi'_{\vec k_2} + \vec k_1\cdot\vec k_2\,\pi_{\vec k_1} \pi_{\vec k_2} \right) \right] \;.
\ee
In the late time limit $\eta\to 0$, we find 
\be
F(\vec k_1, \vec k_2) =  (2\pi)^3 \frac{\kappa}{4\epsilon c_s \mpl^2} \frac{3k_1^2k_2^2+ \vec k_1\cdot\vec k_2(k_1^2 + k_1k_2 +k_2^2)}{4k_1^3k_2^3(k_1+k_2)} \;.
\ee
Plugging back this result in Eq.~\eqref{2pfreal} and integrating by parts we are left with three kind of terms: with two derivatives on $F$, with one derivative on $F$ and one on the exponent, and with two derivatives on the exponent. All of them are multiplied by the delta function and have to be evaluated for $\vec k_2=-\vec k_1$. It can be easily shown that 
\be
F(\vec k_1,-\vec k_1)=0\;, \quad \mathrm{and} \quad \nabla_{k_1}F(\vec k_1, \vec k_2)\Big|_{\vec k_2 = -\vec k_1}=0 \;.
\ee
This means that the only term that survives is the one where two derivatives act on the function $F(\vec k_1, \vec k_2)$. The perturbation of the two-point function is therefore
\be
\delta \langle \pi_{\vec k_1} \pi_{\vec k_2} \rangle = \delta(\vec k_1 + \vec k_2) \nabla^2_{k_{1}}F\Big|_{\vec k_2 = -\vec k_1} =  (2\pi)^3 \delta(\vec k_1 + \vec k_2) P_{\pi}(k_1)\frac{\kappa}{2k^2} \;,
\ee
so that the contribution to the squeezed limit of the three-point function becomes
\be
\langle \zeta_{\vec q} \zeta_{\vec k_1} \zeta_{\vec k_2} \rangle'_{q\to 0} = P_\zeta(q) P_\zeta(k)\frac{q^2}{3k^2} \;.
\ee
If we sum up this result with the contributions \eqref{eq:nogeo}, we get
\be
\langle \zeta_{\vec q} \zeta_{\vec k_1} \zeta_{\vec k_2} \rangle'_{q\to 0} = P_\zeta(q) P_\zeta(k)\left( \frac{5}{6}+\frac{c_3}{2} - \frac{19}{12c_s^2} - \frac{c_3}{2c_s^2} + \frac{3}{4}c_s^2 \right) \frac{q^2}{k^2} \;.
\ee
which agrees with of Eq.~\eqref{squeezed3pf} once the angular average is taken into account. This concludes our check of the consistency relation.

\section{Conclusions}
We have shown and verified in few examples that, in single-field models of inflation, the three-point function in the squeezed limit at order $q^2$ can be understood as the effect of spatial curvature on the short modes, i.e.~calculating the power spectrum in a curved FRW. This gives a nice alternative way to understand the connection between large non-Gaussianity and freezing at a scale shorter than $H$. In some sense this represents the ``last" consistency relation for scalars (for relations including also tensor modes see~\cite{Hinterbichler:2013dpa}): at order $q^3$ one cannot think about non-Gaussianity as the effect of a classical background mode modifying the dynamics of the short modes.

\subsection*{Acknowledgements}
It is a pleasure to thank D. Lopez Nacir and Matias Zaldarriaga for useful comments. Ashley Perko is supported by a Gabilan Stanford Graduate Fellowship.  Leonardo Senatore is supported by DOE Early Career Award DE-FG02-12ER41854 and by NSF grant PHY-1068380.

\appendix

\section{Full diffeomorphism from flat to curved EFTI\label{app:gauge}}
In section \ref{subsec:eftaction} we used a simple physical argument to find the curved speed of sound by giving a VEV to $\pi$. Alternatively it is possible to derive the curved action directly by doing the full diffeomorphism from the flat action (\ref{flat_action}). Using this method the speed of sound can be read off directly from the action in curved coordinates in terms of $c_s$ and $c_3$. We will check that this procedure gives the same result for $\tilde c_s$ as Eq.~(\ref{Eq.curved_cs}). 

To find the action for $\tilde \pi$ we have to transform (\ref{flat_action}) to $\tilde x$ coordinates and then introduce the Goldstone boson $\tilde \pi$. It is important to note that the two Goldstone modes -- $\tilde \pi$ in the curved universe and $\pi$ in the flat universe -- are different given that we are expanding around different backgrounds. As a consequence, in order to go from the action \eqref{flat_action} to the action for the short modes in curved coordinates $\tilde \pi$ we have to do two steps. One is the change of coordinates that brings us to the locally curved universe and the other is a field redefinition that corresponds to the redefinition of $\pi$ due to the change of the background. In order to combine these two steps we have to take $\tilde t \to \tilde t +\tilde \pi(\tilde t)$, where the curved FRW coordinates are related to the flat FRW coordinates by $\tilde t=t+\xi^0(\tilde t)$, with $\xi^0$ given by Eq.~(\ref{xi}). Then we can introduce the curved Goldstone boson $\tilde \pi$ in the action (\ref{flat_action}) using the transformation
\begin{align}
t&=\tilde t+T(\tilde x) \ , \ret
x&=\tilde x +\xi^i(\tilde t) \ ,
\label{transformation}
\end{align}
where $T(\tilde x)=\tilde \pi (\tilde x)(1-\dot \xi^0(\tilde t))-\xi^0(\tilde t)$. Under this change of coordinates the parameters in the action transform as
\begin{align}
g^{00}(x)&=\tilde g^{\mu \nu}(\tilde x)\frac{\partial(\tilde t+T(\tilde x))}{\partial \tilde x^\mu}\frac{\partial(\tilde t+T(\tilde x))}{\partial \tilde x^\nu} \ ,  \  H_0(t)=H_0(\tilde t +T(\tilde x)) \ ,  \ \ldots \ .
\end{align}
Implementing the transformation (\ref{transformation}) on the action (\ref{flat_action}), we find that at quadratic order in $\tilde \pi$ the action for the short modes in curved coordinates is
\begin{align}
\tilde S &=\int \rmd^4\tilde x \   \epsilon    \mpl^2 H_0^2 a_0^3\paren{1+\frac{3}{2} \partial_i\partial_j\zeta_L|_0 \tilde x^i \tilde x^j}  \left[ \frac{1}{c_s^2}\paren{1+\frac{\partial^2 \zeta_L|_0}{2 a_0^2 H_0^2}\paren{-1-4c_3+2\paren{-5+2c_3}c_s^2+6c_s^4}}\dot{\tilde \pi}^2 \right. \ret
&\left. -\paren{1-\partial_i\partial_j\zeta_L|_0 \tilde x^i \tilde x^j}\paren{1 -\frac{\partial^2 \zeta_L|_0}{2a_0^2 H_0^2}\paren{3+2 c_s^2}}\frac{\paren{ \partial_i \tilde \pi}^2}{a_0^2} -\frac{\partial_i\partial_j\zeta_L|_0}{a_0^2 H_0^2}\frac{ \partial_i \tilde \pi \partial_j \tilde \pi}{a_0^2}-2\frac{\partial^2 \zeta_L|_0}{a_0^2}\tilde \pi^2 \right] \ .
\label{anis_action}
\end{align}
After angular averaging we find 
\begin{align}
\tilde S_{\rm avg}&{=} \int \rmd^4\tilde x \    \epsilon   \mpl^2 H_0^2 a_0^3   \paren{1-\frac{3}{4}\kappa \tilde x^2} \left[\frac{1}{ c_s^2}\paren{1+\frac{3\kappa}{4 a_0^2 H_0^2}\paren{1+4 c_3 +2\paren{5-2 c_3}c_s^2-6 c_s^4}}\dot{\tilde \pi}^2 \right. \ret
&\left. -\paren{1+\frac{1}{2}\kappa \tilde x^2}\paren{1+\frac{\kappa}{4a_0^2H_0^2}\paren{7+6 c_s^2}} \frac{\paren{\partial_i \tilde \pi}^2}{a_0^2}+3 \frac{\kappa}{a_0^2}\tilde \pi^2\right] \ .
\label{match_action}
\end{align} 
We can match this to the angular average of the action \eqref{curved_pi}, the most general effective action that has the metric (\ref{curved_eft}), to find $\tilde c_s$. We find that the actions (\ref{match_action}) and \eqref{curved_pi} match if the speed of sound in the curved background is given by Eq.~(\ref{Eq.curved_cs}). Thus we confirm that this is the correct speed of sound in the curved background.

We can use the action (\ref{anis_action}) to compute the fully anisotropic squeezed limit of the three-point function in the same way that we calculated the isotropic limit in Section \ref{subsec.eftfull}.

 The non-geometric two-point function that we find from the equation of motion of Eq. (\ref{anis_action}) can be solved straightforwardly. We find that the late time limit is
\be
\langle \zeta_{\vec k} \zeta_{-\vec k} \rangle'_\kappa = P_\zeta(k)\left[ 1+ \left( -\frac{1}{2} -\frac{c_3}{2} + \frac{2}{c_s^2} + \frac{c_3}{2c_s^2} - \frac{3}{4}c_s^2 \right) \frac{\partial^2\zeta_L|_0}{k^2} - \frac{5}{4c_s^2} \frac{\partial_i\partial_j\zeta_L|_0 k_ik_j}{k^4}\right] \;.
\label{ng_twopoint}
\ee
Correlating Eq. (\ref{ng_twopoint}) with the long mode gives the non-geometric part of the squeezed limit of the three-point function,
\be
\label{squeezed1}
\langle \zeta_{\vec q} \zeta_{\vec k_1} \zeta_{\vec k_2} \rangle'_{q\to 0} = P_\zeta(q) P_\zeta(k) \left[ \left( -\frac{2}{c_s^2} - \frac{c_3}{2c_s^2} + \frac{1}{2} + \frac{c_3}{2} + \frac{ 3 c_s^2}{4} \right) \frac{q^2}{k^2} + \frac{5}{4c_s^2} \frac{(\vec q \cdot \vec k)^2}{k^4} \right] \;.
\ee

Now we need to include geometric effects. The geometric interaction Hamiltonian is given by
\be
H_{int} =\int \rmd^4 x \    \epsilon  \mpl^2 H_0^2 a_0^2     \left[ -\frac{3}{2}\partial_i\partial_j\zeta_L|_0 x^ix^j \frac{1}{c_s^2}{\pi '}^2 + \frac{1}{2}\partial_i\partial_j\zeta_L|_0 x^ix^j  (\partial_a \pi)^2  \right] \; .
\ee
The two-point function can be written as
\begin{align}
\label{2pfreal}
& \int \rmd^3\vec k_1 \rmd^3\vec k_2 \langle \pi_{\vec k_1} \pi_{\vec k_2} \rangle e^{i\vec k_1\cdot\vec x_1 + i\vec k_2\cdot \vec x_2} = \int \rmd^3\vec k_1 \rmd^3\vec k_2 \; F_{ij}(\vec k_1, \vec k_2) \partial_{k_{1i}} \partial_{k_{1j}}\delta(\vec k_1+ \vec k_2) e^{i\vec k_1\cdot\vec x_1 + i\vec k_2\cdot \vec x_2}  \;,
\end{align}
where the function $F_{ij}(\vec k_1,\vec k_2)$ is
\be
F_{ij}(\vec k_1, \vec k_2) = -2(2\pi)^3\; \mathrm{Re} \left[ i \epsilon \mpl^2 \partial_i\partial_j\zeta_L|_0 \; \pi_{\vec k_1}^* \pi_{\vec k_2}^* \int^\eta \frac{\rmd \eta'}{\eta'^2}  \left( \frac{3}{c_s^2}\pi'_{\vec k_1}\pi'_{\vec k_2} + \vec k_1\cdot\vec k_2\pi_{\vec k_1} \pi_{\vec k_2} \right) \right] \;.
\ee
In the late time limit we find 
\be
F_{ij}(\vec k_1, \vec k_2) = -2 (2\pi)^3 \frac{\partial_i\partial_j\zeta_L|_0}{4\epsilon c_s \mpl^2} \frac{3k_1^2k_2^2+ \vec k_1\cdot\vec k_2(k_1^2 + k_1k_2 +k_2^2)}{4k_1^3k_2^3(k_1+k_2)} \;.
\ee
After integrating Eq. (\ref{2pfreal}) by parts we find that the perturbation of the two-point function is
\be
\delta \langle \pi_{\vec k_1} \pi_{\vec k_2} \rangle = \delta(\vec k_1 + \vec k_2) \partial_{k_{1i}}\partial_{k_{1j}} F_{ij}\Big|_{\vec k_2 = -\vec k_1} =  (2\pi)^3 \delta(\vec k_1 + \vec k_2) P_{\pi}(k_1) \frac{5\partial_i\partial_j\zeta_L|_0 k_{1i}k_{1j}- 3\partial^2\zeta_L|_0 k_1^2}{4k_1^4} \;,
\ee
because $F_{ij}$ satisfies
\be
F_{ij}(\vec k_1,-\vec k_1)=0\;, \quad \mathrm{and} \quad \partial_{\vec k_1}F_{ij}(\vec k_1, \vec k_2)\Big|_{\vec k_2 = -\vec k_1}=0 \;,
\ee
just as we found in the isotropic case. The geometric part of the squeezed limit of the three-point function is then
\be
\langle \zeta_{\vec q} \zeta_{\vec k_1} \zeta_{\vec k_2} \rangle'_{q\to 0} = P_\zeta(q) P_\zeta(k) \frac{-5(\vec q \cdot \vec k)^2+ 3q^2k^2}{4k^4} \;.
\ee
Combining this result with the non-geometric contribution \eqref{squeezed1}, we find that the squeezed limit of the three-point function is
\be
\langle \zeta_{\vec q} \zeta_{\vec k_1} \zeta_{\vec k_2} \rangle'_{q\to 0} = P_\zeta(q) P_\zeta(k) \left[ \left( -\frac{2}{c_s^2} - \frac{c_3}{2c_s^2} + \frac{5}{4} + \frac{c_3}{2} + \frac{ 3 c_s^2}{4} \right) \frac{q^2}{k^2} - \frac{5}{4} \frac{(\vec q \cdot \vec k)^2}{k^4}\left( 1-\frac{1}{c_s^2} \right) \right] \;.
\ee
which agrees with Eq.~\eqref{squeezed3pf}. 

\section{K-inflation}\label{app:k-infl}
In this appendix we want to show how it is possible to obtain the same result obtained in section \ref{subsec.eftsmallcs} with the dictionary  of k-inflation. Starting from the action for k-inflation in the curved universe
\begin{equation}	\label{Eq.pxaction}
\begin{split}
S=\int \rmd^4x\sqrt{-g}P(X,\phi) \ ,
\end{split}
\end{equation}
where $X=-\frac{1}{2} g^{\mu\nu}\partial_\mu\phi\partial_\nu\phi$, and expanding around $\phi(t,x)=\phi(t)+\dot \phi(t) \pi(t,x)$, we get the action for the homogeneous background $\phi(t)$
\begin{equation}	
\begin{split}
S=\frac{1}{2}\int \rmd^4x \, \Big(a^3 P_X  \dot\phi^2 -a^3 V(\phi)\Big) \;,
\end{split}
\end{equation}
and the equation of motion
\begin{equation}	\label{Eq.eomback}
\begin{split}
\partial_t (a^3 P_X \dot \phi)-a^3 V'=0 \ .
\end{split}
\end{equation}
Eq.~(\ref{Eq.eomback}) implies that the curvature not only changes the scale factor through Friedmann's equation and the spatial part of the metric but it also affects the background velocity $\dot \phi(t)$. Solving perturbatively for the background field $\phi(t)=\phi_0(t)+\delta \phi (t)$, one obtains\footnote{The homogeneous solution $C a_0^{-3}$ decays faster than the particular solution and thus can be neglected.}
\be
\delta \dot \phi = C \; a_0^{-3} + \frac{3}{2} c_s^2 \dot \phi_0 \frac{\kappa}{a_0^2 H_0^2} \;.
\ee
Thus in $P(X)$ models one has to consider only the effects induced by the change of the background cosmology, which means the variations of $a(t)$, $H(t)$, and $\dot \phi(t)$, and the geometrical effect proportional to $x^2$. To relate the coefficients of the curved action for perturbations,
\be\label{Eq.pxactionpi}
S=\frac{1}{2}\int d^4x\, \sqrt{-g} \Big(  (X P_X  +2X^2 P_{XX}) \dot\pi^2 -X P_X  g^{ij} \partial_i \pi \partial_j\pi + 3\kappa X P_X  \pi^2\Big) \;,
\ee
with those of the flat one, one has to expand Eq.~(\ref{Eq.pxaction}) around flat quantities using the variations 
\be\label{Eq.variations}
 \frac{\delta a(\eta)}{a(\eta)} = \frac{\kappa}{6}\eta^2  \quad \text{and} \quad \frac{\delta X}{X} = 2\frac{\delta \dot \phi}{\dot \phi} = 3 \frac{c_s^2 \kappa}{H_0^2 a_0^2} \;.
\ee
so for example 
\be
P_X \rightarrow P_X \( 1 + \frac{XP_{XX}}{P_X} \frac{\delta X}{X} \) \;, \quad P_{XX} \rightarrow P_{XX} \( 1 + \frac{XP_{XXX}}{P_{XX}} \frac{\delta X}{X} \) \;.
\ee
Using the definition for the parameters $c_s^2$, $\lambda$ and $\Sigma$ of Ref. \cite{Chen:2006nt},  
\be
\frac{XP_{XX}}{P_X} = \frac{1}{2}\( \frac{1}{c_s^2} - 1 \)\ ,  \qquad \frac{2X^2P_{XXX}}{P_X} = \frac{3\lambda}{c_s^2\Sigma} - \frac{3}{2}\( \frac{1}{c_s^2} -1 \) \ ,
\ee
the action in Eq.~(\ref{Eq.pxactionpi}) reads
\begin{align}
S &=\int \rmd^4 x \    H_0^2 \epsilon a_0^2    \left[\frac{1}{c_s^2} \paren{1+ \frac{ \kappa}{3a_0^2 H_0^2} +\frac{9c_s^2\kappa}{a_0^2 H_0^2}\frac{\lambda}{\Sigma}}  {\pi '}^2  -\paren{1+\frac{11\kappa}{6a_0^2H_0^2}} (\partial_i \pi)^2\right] \ .
\end{align} 
Here again we are considering only terms which contribute at order $1/c_s^2$. This greatly simplifies the calculation since the metric can be taken to be spatially flat and the field can be decomposed in Fourier modes. The equation of motion derived from this action is 
\be
\pi ''_{\vec k} - \frac{2}{\eta}\pi '_{\vec k} + c_s^2k^2 \pi_{\vec k} + \frac{\kappa}{ a_0^2  H_0^2} \left(\left(\frac{1}{3}+ 9c_s^2 \frac{\lambda}{\Sigma}\right){\pi ''_{\vec k}} + \frac{11}{6} c_s^2 k^2 \pi_{\vec k} \right)= 0 \;,
\ee
and can easily be solved perturbatively in $\kappa$. Once one has checked that the solution has the correct normalization, the power spectrum for $\zeta$ can be calculated using the standard procedure and the result is
\be
\langle \zeta_{\vec k} \zeta_{-\vec k} \rangle'_\kappa = P_\zeta(k)\left( 1- \frac{19+18\frac{\lambda}{\Sigma}}{8c_s^2k^2}\kappa \right) \;.
\ee
The contribution to the squeezed limit of the three-point function is then obtained by averaging over the long mode and substituting $\kappa=\frac 23 q^2\zeta_{\vec q}$
\be
\langle \zeta_{\vec q} \zeta_{\vec k_1} \zeta_{\vec k_2} \rangle'_{q\to 0} = - P_\zeta(q) P_\zeta(k) \frac{1}{c_s^2} \left(\frac{19}{12}+\frac{3}{2}\frac{\lambda}{\Sigma} \right) \frac{q^2}{k^2}\;.
\ee
Indeed, this agrees with the squeezed limit of the three-point function calculated with the standard \emph{in-in} formalism (the detailed calculation can be found in Ref. \cite{Chen:2006nt}), once the angular average is taken and terms not enhanced by $1/c_s^2$ are neglected. 

\section{Ghost Inflation} \label{sec:ghost}
In this appendix we show that the argument presented in the main part of the paper can be applied also in the case of ghost inflation \cite{ArkaniHamed:2003uz}. The ghost condensate can be thought as a model in which the background field is brought dynamically in a minimum of $P(X)$. This means that the equation of motion for the background 
\be
\partial_t (a^3 P_X \dot \phi)=0 \;
\ee
is trivially satisfied for a constant velocity $\dot \phi$ equal to the unperturbed one, so that the curvature does not change the background solution. The curvature thus will enter only through the change of the scale factor\footnote{We are neglecting here the geometrical effects coming from the $x$-dependent part of the metric.}.
The quadratic action for perturbations in conformal time reads
\begin{equation}	
\begin{split}
S=\int \rmd^4x \,  \left(2 M_2^4 a^2 \pi'^2-\frac{\bar M^2}{2} (\partial^2\pi)^2\right) \ .
\end{split}
\end{equation}
From this, the variation of the two point function in the presence of curvature is easily obtained by solving perturbatively the equation of motion
\begin{equation}	\label{Eq.eomghost}
\begin{split}
4 M_2^4  \partial_\eta\left(a^2_0\left(1+2\frac{\delta a}{a}\right) \pi'\right)  +\bar M^2 k^4\pi=0 \ ,
\end{split}
\end{equation}
by defining $\pi=\pi_0 + \kappa H_0^{-2}\pi_1$. The solution of the unperturbed equation
\begin{equation}	
\begin{split}
4 M_2^4     \pi_0''+8M_2^4 \mathcal{H}_0   \pi_0' +\bar M^2 \frac{k^4}{a^2_0}\pi_0=0 \ ,
\end{split}
\end{equation}
is given by \cite{ArkaniHamed:2003uz}
\begin{equation}\label{Eq.pi}	
\begin{split}
\pi_0(k,\eta)=2^{-1/4}H_0^{1/4}M_2^{-1/2}\bar M^{-3/4}k^{-3/2}F\left(\frac{H_0^{1/2}\bar M^{1/2}}{\sqrt{2}M_2}k\eta\right) \ ,
\end{split}
\end{equation}
where we defined $F(x)=\sqrt{\pi/8} (-x)^{3/2} H^{(1)}_{3/4}(x^2/2)$, and $H^{(1)}_{\alpha}$ is the Hankel function of first kind. After expanding  Eq.~(\ref{Eq.eomghost}) at first order in $\kappa$, we have to solve  
\begin{equation}	
\begin{split}
4 M_2^4     \pi_1''+8M_2^4 \mathcal{H}_0   \pi_1' +\bar M^2 \frac{k^4}{a^2_0}\pi_1=-\frac{4 M_2^4}{3a^2_0}\pi_0'' \;.
\end{split}
\end{equation}
This can be done using the Green function of the unperturbed equation of motion, which is given by the commutator of the free fields
\begin{equation}	
\begin{split}
G(x_1,x_2)=i \,a^{-2}(\eta_2)\, \theta(\eta_1-\eta_2)\,[\pi_0(x_1),\pi_0(x_2)] \;.
\end{split}
\end{equation}
The first order solution is then given by
\begin{equation}	
\begin{split}
\pi_1(k_1,\eta)&=-\frac{4 H_0^4 M_2^4}{3} \int_{-\infty}^{\eta} \rmd\tau \,a_0^4(\tau)\, G(k_1,\eta;k_2,\tau)\,\tau^2\pi_0''(k_2,\tau)\\
&=-\frac{4 H_0^2M_2^4}{3} \int_{-\infty}^{\eta} \rmd\tau \, i\theta(\eta-\tau)\,[\pi_0(k_1,\eta), \pi_0(k_2,\tau)]\,\pi_0''(k_2,\tau) \;.
\end{split}
\end{equation}

The field $\pi_0$, being the solution of the unperturbed equation of motion, is normalized in such a way that
\begin{equation}	\label{Eq.norm0}
\begin{split}
[\pi_0(\eta),\Pi_{\pi_0} (\eta)]_{\kappa=0}=4M_2^4 a_0^2[\pi_0(\eta),\pi'_0(\eta)]=i \;,
\end{split}
\end{equation}
where $\Pi_{\pi_0}$ is the conjugated momentum. We can check  that the same condition 
is satisfied by $\pi$ at first order in $\kappa$ without changing the normalization of the homogeneous part, by direct evaluation of the  commutator
\begin{equation}	\label{Eq.comm}	
\begin{split}
[\pi(\eta),\Pi_{\pi} (\eta)]&=4M_2^4 a^2[\pi_0(\eta),\pi'_0(\eta)]+4M_2^4 a_0^2\kappa H_0^{-2}\big([\pi_0(\eta),\pi'_1(\eta)]+[\pi_1(\eta),\pi'_0(\eta)]\big)\\
&=i+ i\frac{\kappa\eta^2}{3}+4M_2^4 a_0^2\frac{\kappa}{H_0^2}\big([\pi_0(\eta),\pi'_1(\eta)]+[\pi_1(\eta),\pi'_0(\eta)]\big) \ ,
\end{split}
\end{equation}
where the commutators involving $\pi_0$ and $\pi_1$ are 
\begin{equation}	
\begin{split}
[\pi_0(\eta),\pi'_1(\eta)]&=-\frac{4 H_0^2M_2^4}{3} \int_{-\infty}^{\eta} \rmd\tau \, i\theta(\eta-\tau)\,[\pi_0'(\eta), \pi_0(\tau)]\,[\pi_0(\eta),\pi_0''(\tau)] \;,\\
[\pi_1(\eta),\pi'_0(\eta)]&=-\frac{4 H_0^2M_2^4}{3} \int_{-\infty}^{\eta} \rmd\tau \, i\theta(\eta-\tau)\,[\pi_0(\eta), \pi_0(\tau)]\,[\pi_0''(\tau),\pi_0'(\eta)] \ .
\end{split}
\end{equation}
After many integrations by parts the commutators appearing in the RHS can be written as commutators at equal time.
Furthermore, when the two are summed we obtain
\begin{equation}	
\begin{split}
[\pi_0(\eta),\pi'_1(\eta)]+ [\pi_1(\eta),\pi'_0(\eta)]&=-i\frac{4 M_2^4}{3}[\pi_0(\eta),\pi_0'(\eta)][\pi_0'(\eta),\pi_0(\eta)]=-i\frac{1}{12 M_2^4 a_0^4} \ .
\end{split}
\end{equation}
Plugging this result back in Eq.~(\ref{Eq.comm}) gives $[\pi(\eta),\Pi_{\pi} (\eta)]=i$, thus the solution is correctly normalized.\\

The  two point function in the presence of the long mode is then given by the following expression 
\begin{equation}	
\begin{split}
\langle \pi\pi\rangle'_{\pi_L}&=\pi^*(0)\pi(0)=P_\pi+ \frac{2 \kappa}{H_0^2}\, \text{Re}[\pi_0^*(0)\pi_1(0)] \ ,
\end{split}
\end{equation}
where $P_\pi(k)=\langle \pi_0\pi_0\rangle'=2^{-1/2}H_0^{1/2}M_2^{-1}\bar M^{-3/2}|F(0)|^2 k^{-3}$, 
and its perturbation  is
\begin{equation}	
\begin{split}
\delta \langle \pi\pi\rangle'_{\pi_L}=\frac{2 \kappa}{H_0^2}\, \text{Re}[\pi_0^*(0)\pi_1(0)]&=-\frac{4 M_2^2\kappa }{3H_0 \bar M^{3}k^{5}}\frac{H_0^{1/2}\bar M^{1/2}}{\sqrt{2}M_2} \, \text{Re}\left[i F^*(0)\int_{-\infty}^0 \rmd x \big(F(0)F^*(x)-F^*(0)F(x)\big)F''(x) \right]\\
&=-\frac{8 M_2^2\kappa }{3H_0 \bar M^{3}k^{5}}\frac{H_0^{1/2}\bar M^{1/2}}{\sqrt{2}M_2} |F(0)|^2\, \text{Re}\left[i\int_{-\infty}^0 \rmd x \, \text{Re}[F(x)]F''(x) \right]\\
&=\frac{8 M_2^2\kappa }{3H_0 \bar Mk^{2}} P_\pi(k) \, \text{Im}\left[\int_{-\infty}^0 \rmd x \, \text{Re}[F(x)]F''(x) \right] \;.
\end{split}
\end{equation}
Now we can use the following property of the function $F(x)$
\begin{equation}	\label{Eq.reim}
\begin{split}
\int_{-\infty}^0 \rmd x \, \text{Re}[F(x)]F''(x)=\frac{1}{2}\int_{-\infty}^0 \rmd x \, F(x)F''(x) \;,
\end{split}
\end{equation}
to rewrite the variation of the power spectrum as 
\be
\delta \langle \pi_{\vec k} \pi_{-\vec k} \rangle'_{\pi_L} = \frac{4 M_2^2\kappa }{3H_0 \bar Mk^{2}} P_\pi(k) \, \text{Im}\left[\int_{-\infty}^0 \rmd x \, F(x)F''(x) \right] \;.
\ee
Taking the average over the long mode and using $\zeta=-H_0\pi$ and $\kappa=\frac{2}{3}q^2 \zeta_{\vec q}$, we find the three-point function in the squeezed limit
\be \label{Eq.3pfghost}
\langle \pi_{\vec q} \pi_{\vec k_1} \pi_{\vec k_2} \rangle'_{q \to 0} = -\frac{8 M_2^2 }{9 \bar M} P_\pi(q) P_\pi(k) \frac{q^2 }{k^2} \text{Im}\left[\int_{-\infty}^0 \rmd x \, F(x)F''(x) \right] \;.
\ee

This result has to be compared with the squeezed limit of the three-point function in \cite{ArkaniHamed:2003uz},
\be
\langle \pi_{\vec k_1} \pi_{\vec k_2} \pi_{\vec k_3} \rangle' = N \frac{1}{\prod k_a^3} 2\mbox{Re} \left[ \int_{-\infty}^0 \frac{\rmd \eta}{\eta}\( F^*(k_1\eta) F^*(k_2 \eta) F'^* (k_3\eta) k_3(\vec k_1 \cdot \vec k_2) + \mbox{symm.} \)  \right] \;,
\ee
where the constant $N$ is given by
\be
N= -\frac{(2\pi)^{3/2}}{\Gamma(1/4)^3} \frac{H_0^5}{2M_2^2} \( \frac{2M_2^2}{\bar M H_0} \)^4 \( \frac{1}{2M_2^2} \)^3 \;.
\ee
The terms in the parentheses are proportional at least to $q$, so if we use the symmetric expansion \eqref{symexp} the product in front of the integral gives only the leading contribution $1/q^3k^6$ (corrections are of order $q^2$). Keeping only the terms contributing at order $q^2$, we get
\be
\langle \pi_{\vec q} \pi_{\vec k_1} \pi_{\vec k_2} \rangle'_{q\to 0} = -\frac{8}{3} \frac{M_2^2}{\bar M} P_\pi(q)P_\pi(k) \frac{q^2}{k^2} \mbox{Im} \left[  \int_{-\infty}^0 \rmd x \( -\frac{1}{x} F^*(x)F'^*(x) + F^*(x) F''^*(x)\) \right]\;.
\ee
We can then numerically show that the following relation holds
\be
\mbox{Im}   \int_{-\infty}^0 \rmd x \( -\frac{1}{x} F^*(x)F'^*(x) \) = - \frac{4}{3}\; \mbox{Im}   \int_{-\infty}^0 \rmd x  F^*(x) F''^*(x) \;,
\ee
such that
\be
\mbox{Im} \left[  \int_{-\infty}^0 \rmd x \( -\frac{1}{x} F^*(x)F'^*(x) + F^*(x) F''^*(x)\) \right] = \frac{1}{3}\; \mbox{Im}   \int_{-\infty}^0 \rmd x  F(x) F''(x) \;.
\ee
Finally, we get for the squeezed limit of the three-point function
\be
\langle \pi_{\vec q} \pi_{\vec k_1} \pi_{\vec k_2} \rangle'_{q\to 0} = -\frac{8}{9} \frac{M_2^2}{\bar M} P_\pi(q)P_\pi(k) \frac{q^2}{k^2} \mbox{Im}   \int_{-\infty}^0 \rmd x  F(x) F''(x)  \ , 
\ee
which indeed agrees with Eq.~(\ref{Eq.3pfghost}).

\section{Khronon Inflation} \label{sec:khronon}
In this appendix we want to check that our consistency relation \eqref{eq:schematic} works also for Khronon Inflation \cite{Creminelli:2012xb}. We concentrate only on terms enhanced by $c_s^{-2}$ and on the isotropic case. The action for Khronon Inflation at leading order in derivatives and in the decoupling limit is
\begin{equation}	
\begin{split}
S=\frac{1}{2}\int \rmd^4x\sqrt{-g}  \left(M_\lambda^2 (\nabla_\mu u^\mu +3H_0)^2+M_\alpha^2 u^\mu u^\nu \nabla_\mu u_\rho \nabla_\nu u^\rho \right) \ ,
\end{split}
\end{equation}
where $u_\mu=\frac{\partial_\mu \phi}{ \sqrt{g^{\alpha\beta}\partial_\alpha \phi\partial_\beta \phi}}$.  Since in this model time reparametrization is promoted to an exact symmetry, the curvature will not affect the background solution and perturbations can be expanded around $\phi(t)=t+\pi$. Moreover the term proportional to $M_\alpha^2$, when expanded at quadratic order, does not depend on the scale factor $a$: since this is the only term enhanced by $c_s^{-2}=(M_\lambda/M_\alpha)^{-2}$, we expect no variation of the two-point function at this order. Indeed the action   for perturbations, at first order in curvature, reads
\begin{equation}	
\begin{split}
S=\frac{1}{2}\int \rmd^4x  \left(M_\alpha^2 (\partial \pi')^2-M_\lambda^2 \left((\partial^2 \pi)^2-6a^2(H_0-H) H (\partial \pi)^2\right) \right) \;.
\end{split}
\end{equation}
Expanding the equation of motion at first order in $\kappa$, we see that the curvature does not source any variation of the action enhanced by $c_s^{-2}$, so there is no variation of the two-point function $\propto c_s^{-2}$. This means that the three-point function in the squeezed limit at order $q^2$, after we take the angular average, should not be enhanced by $c_s^{-2}$. Indeed the three-point function $\propto c_s^{-2}$ reads \cite{Creminelli:2012xb}
\be
\langle \zeta_{\vec k_1} \zeta_{\vec k_2} \zeta_{\vec k_3} \rangle' = \frac{1}{\prod k_i^3} P_\zeta^2\bigg[ - \frac{1}{c_s^2}\frac{k_1^3}{k_t^2}\vec{k}_2\cdot\vec{k}_3\bigg] +\, {\rm cyclic \;perms.}\,,
\ee
and the angular average of the $\mathcal O(q^2)$ term is zero. The consistency relations holds.

\end{document}